\documentclass{llncs}
\usepackage{times}
\usepackage{url}
\usepackage{dsfont}
\usepackage{color}
\usepackage{graphicx}
\usepackage{mathtools}
\usepackage{amssymb}
\usepackage{algorithm}
\usepackage[noend]{algorithmic}
\usepackage{array}
\usepackage[utf8]{inputenc}
\usepackage{enumerate}

\newcommand{\argmin}{\arg\!\min}
\begin{document}

\title{\textsc{Fixed-Parameter Algorithms for Rectilinear Steiner tree and Rectilinear Traveling Salesman Problem in the plane}}
\author{Hadrien Cambazard \and Nicolas Catusse\thanks{Nicolas Catusse is partially supported by LabEx PERSYVAL-Lab (\textsc{anr-11-labx-0025}) and a grant PEPS JCJC from CNRS, entitled OTARI.}}
\institute{Univ. Grenoble Alpes, G-SCOP, F-38000 Grenoble, France \\
CNRS, G-SCOP, F-38000 Grenoble, France \\
\email{\{hadrien.cambazard|nicolas.catusse\}@grenoble-inp.fr}}
\maketitle

\begin{abstract}
Given a set $P$ of $n$ points with their pairwise distances, the traveling salesman problem (TSP) asks for a shortest tour that visits each point exactly once.
A TSP instance is rectilinear when the points lie in the plane and the distance considered between two points is the $l_1$ distance. In this paper, a fixed-parameter algorithm for the Rectilinear TSP is presented and relies on techniques for solving TSP on bounded-treewidth graphs. It proves that the problem can be solved in $O\left(nh7^h\right)$ where $h \leq n$ denotes the number of horizontal lines containing the points of $P$. The same technique can be directly applied to the problem of finding a shortest rectilinear Steiner tree that interconnects the points of $P$ providing a $O\left(nh5^h\right)$ time complexity. Both bounds improve over the best time bounds known for these problems.
\end{abstract}


\section{Introduction}
\label{intro}
Given a set $P$ of $n$ points with their pairwise distances, the Traveling Salesman Problem (TSP) asks for a shortest tour that visits each point exactly once.
We consider the Rectilinear instances of the TSP (RTSP), \emph{i.e.} TSP instances where the points lie in the plane and the distance between two points is the sum of the differences of their x- and y-coordinates. This metric is commonly known as the $l_1$ distance, the Manhattan distance or the city-block metric. We are also interested in the rectilinear variant of the minimum Steiner tree problem (RST). The RST problem is to find a shortest Steiner tree that interconnects the points of $P$ using vertical and horizontal lines. A number of studies have considered the rectilinear Steiner tree in the past, motivated by the applications to wire printed circuits.
Let $h$ (resp. $v$) be the number of horizontal (resp. vertical) straight lines that contain the points of $P$. Note that without loss of generality, we can assume that $h\leq v$. We present fixed-parameter algorithms using $h$ as a parameter for RTSP and RST. The RTSP is expressed as a Steiner (also called Subset) TSP in a planar grid-graph of pathwidth equal to $h$. Our approach is then based on standard techniques for solving TSP and Steiner tree on bounded-treewidth graphs \cite{Bodlaender1993,Cook2003}. A number of existing results exploiting planarity or bounded treewidth/pathwidth can thus be applied and will be reviewed in this introduction. \\

The contribution of this paper is a refined complexity analysis showing that RTSP and RST can be solved respectively in time $O\left(nh7^h\right)$ and $O\left(nh5^h\right)$. A specific structure called \emph{non-crossing partitions} lies at the core of the complexity of problems that involve a global connectivity constraint in planar graphs \cite{Aroraetal98,Dorn2010,Dorn2012}. Moreover, the number of such non-crossing partitions relates to Catalan numbers which have been used in the past to establish complexity results. 

The $O\left(nh7^h\right)$ and $O\left(nh5^h\right)$ runtimes improve over the bounds known from dedicated approaches \cite{Brazil2000,Fomin2016}, dynamic programming algorithms dedicated to weighted problems in planar graphs \cite{Dorn2010,Klein2014} or the recent and more generic framework of \cite{Bodlaender2015}.
The algorithm proposed for RTSP generalizes the approaches designed for specific placement of the points in the context of warehouses \cite{rat83,roo01}. The parameter is meaningful in a number of routing applications and the resulting time complexity is low enough for the algorithm to be used in practice as demonstrated by the experimental evaluation.\\

Firstly, the existing results about RTSP will be reviewed. The NP-completeness proof by Itai, Papadimitriou, and Szwarcfiter \cite{ita82} for the Euclidean TSP immediately implies the NP-completeness of the following type of TSP: points are in $\mathbb{R}^2$ and distances are measured according to a polyhedral norm. In particular, the TSP with Manhattan distances is NP-complete. A similar parameter to $h$ has been used by Rote \cite{rot92}. When the points of $P$ lie on a small number $h$ of parallel lines in the plane, Rote shows that the Euclidean TSP problem can be solved with a dynamic programming approach in time $O(n^h)$. Moreover, this algorithm can be applied with the $l_1$ metric. Note also that a variant of the RTSP where the objective is to minimize the number of bends has been widely studied and in particular using fixed-parameter tractability \cite{est10}.  A related case of the RTSP is considered by Ratliff and Rosenthal \cite{rat83} to compute the shortest order picking tour in a warehouse with rectangular layout. Warehouses are usually designed with multiple parallel vertical aisles and multiple horizontal cross aisles. The algorithm of \cite{rat83} solves the problem when the warehouse does not have any cross aisles (\emph{i.e.} a case similar to $h=2$) and Roodbergen and De Koster \cite{roo01} extend this algorithm to handle the case of one single cross aisle (\emph{i.e.} a case similar to $h=3$). Beyond the RTSP, many results related to Euclidean or Subset/Steiner TSP are relevant to this paper. The Polynomial-Time Approximation Scheme (PTAS) for the Euclidean TSP, proposed by Arora \cite{aro98}, can be applied with any geometric norm, such as $l_p$ for $p \ge 1$ or other Minkowski norms. It provides, for any $c > 1$, a $\left(1+\frac{1}{c}\right)$-approximation running in time $O(n(\log n)^{O(c)})$.
A number of subexponential exact algorithms of running time $2^{O(\sqrt{n}log(n))}$ have been designed independently for the Euclidean TSP by taking advantage of planar separators \cite{Hwang1993}. A recent result inheriting from this line of research is a $O(2^{9.8594\sqrt{n}})$ for the TSP on planar graphs proposed by Dorn et al. \cite{Dorn2010} and exploiting non-crossing partitions. Such partitions are at the heart of the present paper and have been used already by Arora et al. \cite{Aroraetal98} to achieve better approximation schemes. The link between non-crossing partitions and the Catalan number has been investigated further by Dorn et al. \cite{Dorn2012} when applying dynamic programming to H-minor-free graphs. A recent parameterized algorithm was also proposed in \cite{Klein2014} for the Subset TSP (also referred to as Steiner TSP) on planar graphs with integer weights no greater than a constant W. The problem is to find the shortest closed walk visiting all vertices of a subset of $k$ vertices. It is solved by \cite{Klein2014} in $(2^{\sqrt{k} log(k)} + W)n^{O(1)}$. Since we recast the RTSP as a Steiner TSP in a graph of $hv$ vertices where a subset of $n$ of them must be visited, the result of \cite{Klein2014} yields a $(2^{\sqrt{n} log(n)} + W)(hv)^{O(1)}$ algorithm which is more general but incomparable to the complexity proposed in the present paper.

The Steiner tree problem has been extensively studied and exact algorithms for the rectilinear variant have been proposed. An exact approach for $h=2$ is proposed by \cite{aho1977} in 1977 and can be generalized for higher values of $h$. It can be shown that the resulting algorithm have a time complexity of $O(n16^h)$ (see \cite{Brazil2000}). The algorithm is implemented by Ganley and Cohoon \cite{Ganley1996} but the high time complexity limits its applicability. A second exact approach based on dynamic programming is later proposed by \cite{Brazil2000} with a $O(nh^310^h)$ time complexity. Arora also proposed a PTAS for the Steiner tree problem in his paper dedicated to TSP \cite{aro98}. Similarly to TSP, a subexponential algorithm with a $2^{O(\sqrt{n}log(n))}$ runtime has been recently designed \cite{Fomin2016}.

It also worth mentioning the work of Bodlaender et al. \cite{Bodlaender2015} which addresses a wide range of graph problems with bounded treewidth $tw$ and a global connectivity property such as the Hamiltonian Cycle, Steiner tree or Traveling Salesman problems. It was not known for a long time how to improve the $tw^{tw}n^{O(1)}$ time complexity of the known dynamic programming approaches taking advantage of a bounded treewidth $tw$ of the underlying graph. The \emph{rank based approach} proposed in \cite{Bodlaender2015} however provides algorithms with a $c^{tw}n^{O(1)}$ time complexity, where $c$ is a small constant, for both weighted and counting versions of these problems. In particular, the Steiner Tree can be solved in $n(1+2^{w})^{h}h^{O(1)}$ where $w$ is the matrix multiplication exponent (the best known upper bound for $w$ is currently $2.3727$ \cite{Williams2012}). We believe that the use of the rank based approach for the Steiner TSP would lead to a $n(1 + 2^{w+1})^{h} h^{O(1)}$ runtime (the details are in Section \ref{rankbased}) and that the improvement obtained by \cite{Cygan2013} for TSP is not directly applicable. Both runtimes  thus remain improved by our dedicated analysis. 

In section \ref{algotsp}, we present a dynamic programming approach for the RTSP by describing the possible states, transitions, and the main algorithm (Section \ref{secalgo}) whereas Section \ref{complexity} deals with the complexity analysis of the algorithm. Section \ref{algosteiner} follows the same outline for the RST. Section \ref{rankbased} is focused on the comparison with the rank based technique.
Some experimental results for both problems are presented in Section \ref{expe} followed by concluding remarks in section \ref{conc}.

\section{A Fixed-parameter algorithm for the Rectilinear TSP}
\label{algotsp}
\begin{figure}[b]
   \begin{minipage}[c]{.49\linewidth}
      	\includegraphics[scale=.75]{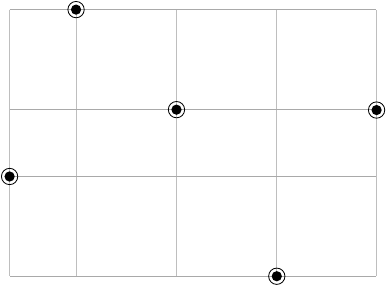}
	\centering
	\caption{The Hanan grid $\Gamma(P)$}
	\label{figHanan}
   \end{minipage} \hfill
   \begin{minipage}[c]{.49\linewidth}
      	\includegraphics[scale=.75]{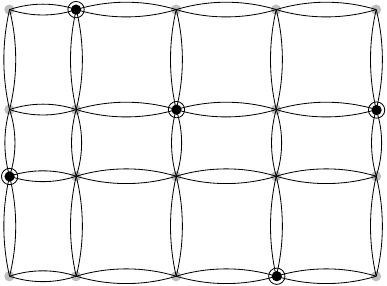}
	\centering
	\caption{The graph $G$}
	\label{figG}
   \end{minipage}
\end{figure}

As above, let $P$ denote the given set of points in the plane.
The \textbf{Hanan grid} $\Gamma(P)$ is the set of segments obtained by constructing vertical and horizontal lines through each point of $P$. Recall that $h$ and $v$ denote the number of horizontal and vertical lines of $\Gamma(P)$. Fig.~\ref{figHanan} shows an instance where $|P| = 5$ and the corresponding Hanan grid $\Gamma(P)$. Note that, for all the figures, the points of $P$ are indicated by circled black dots.

Let $l_1(p_1, p_2)$ be the $l_1$ distance between two points $p_1$ and $p_2$.
Let define the \textbf{undirected graph} $G = (V, E)$ by associating a vertex to each intersection of $\Gamma(P)$ and two parallel edges for each segment of $\Gamma(P)$, with length equal to the $l_1$ distance between the intersections (see Fig.~\ref{figG}). We denote by $v_{i,j}$ with $i \in [1, h]$ and $j \in [1, v]$ the vertex at the intersection between horizontal line $i$ and vertical line $j$. The points of $P$ are thus related to some of the vertices of $G$ and we have $P \subseteq V$. Moreover, $L= (V_L, E_L)$ is said to be a subgraph of $G$ (denoted $L \subseteq G$) when $V_L \subseteq V$ and $E_L \subseteq E$. Finally, the degree of a vertex $v \in V$ is the number of edges in $E$ incident to $v$.

The problem is to find a shortest tour in $G$, visiting all points of $P$. The TSP problem is thus solved as a Steiner TSP problem on $G$. The Steiner TSP is a variant of TSP where the graph is not complete, only a subset of the vertices must be visited by the salesman, vertices may be visited more than once and edges may be traversed more than once \cite{Letchford201383}. Any optimal tour in $G$ visiting all points of $P$ \emph{i.e.} an optimal solution of the Steiner TSP problem where $P$ is the set of mandatory vertices, is also an optimal solution of the rectilinear TSP problem. The points (vertices) of $V-P$ can be used to change direction and can lie on a rectilinear path connecting two vertices of $P$. We will show that $G$ always contains an optimal solution of the original rectilinear TSP problem.

\begin{figure}[b]
   \begin{minipage}[c]{.49\linewidth}
	\centering
	\includegraphics[scale=.80]{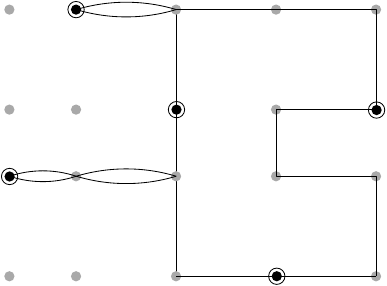}
	\caption{A tour subgraph}
	\label{figTourSubgraph}
   \end{minipage} \hfill
   \begin{minipage}[c]{.49\linewidth}
      	\centering
	\includegraphics[scale=.80]{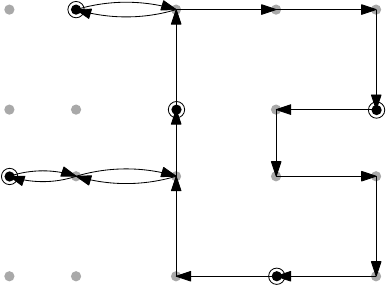}
	\caption{A directed tour subgraph}
	\label{figTourSubgraphOrient}
   \end{minipage}
\end{figure}

Notice that we do not need to consider a directed graph, \emph{i.e.} $G$ is undirected, because the algorithm builds a shortest \emph{tour subgraph} that can be directed as a post-processing step. A subgraph $T$ of $G$ that contains all points $P$ will be called \textbf{a tour subgraph} if there is an orientation of the edges that is a tour in which every edge of $T$ is used exactly once (an order-picking tour in \cite{rat83}). Figs.~\ref{figTourSubgraph} and~\ref{figTourSubgraphOrient} show a tour subgraph and a possible orientation. 
The following characterization of a tour subgraph, given in \cite{rat83}, is a specialization of a well known theorem on Eulerian graphs (e.g., see Christofides \cite{chr75}):

\begin{theorem}(adapted from \cite{rat83}) \label{th1} A subtour $T \subseteq G$ is a tour subgraph if and only if:
\begin{enumerate}[(a)]
\item All vertices of $P$ belong to the vertices of $T$;
\item $T$ is connected;
\item every vertex in $T$ has an even degree.
\end{enumerate}
\end{theorem}

The tour subgraph of Fig.~\ref{figTourSubgraph} is connected and has 3 vertices of degree 4. Note that parallel edges can be used in a tour subgraph.
From any subgraph that is known to be a tour subgraph, we can easily determine an oriented TSP-tour by using any algorithm finding an Eulerian cycle. Therefore, we will focus on finding a minimum length tour subgraph. Let's now show that an optimal solution of the Steiner TSP problem in $G$ provides an optimal solution of the original RTSP problem.

\begin{lemma} 
Let $l^{*}$ be the length of an optimal solution for the RTSP problem. There exists an optimal tour subgraph in G of length $l^{*}$.
\end{lemma}
\proof
Consider an optimal ordering of $P$ (giving a solution of distance $l^*$) and a directed tour $S$ in $G$ obtained by following a shortest path in $G$ between consecutive points of the ordering. 
We can build a tour subgraph $T$ by adding an edge $(i, j)$ for each arc of $S$. Requirements (a), (b), (c) of Theorem \ref{th1} are satisfied because $S$ is a tour. An edge $(i, j)$ is added at most twice because $S$ is optimal, so $T \subseteq G$.
Finally, it has a length equals to $l^*$ since $G$ contains a shortest rectilinear path between each pair of points. $\hfill \square$
\endproof

The algorithm builds a tour subgraph by completing a \textbf{partial tour subgraph}.

\begin{figure}[t!]
\begin{center}
\includegraphics[scale=0.8]{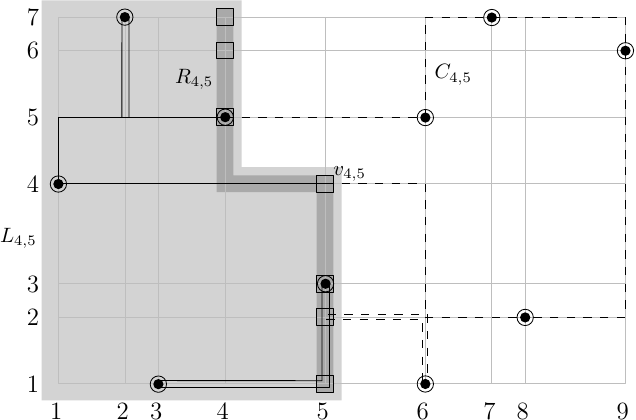}
\caption{Hanan grid $\Gamma(P)$ with $L_{4,5}$ inside the light gray area and $R_{4,5}$ inside the dark gray area. The vertices of $R_{4,5}$ are represented by squares. The black edges represent an $L_{4,5}$ partial tour subgraph and the dashed edges represent one possible completion.}
\label{hanan}
\end{center}
\end{figure}

\begin{definition}\label{d1}
For any subgraph $L \subseteq G$; a subgraph $T \subseteq L$ is an $L$ \textbf{partial tour subgraph} if there exists a subgraph $F \subseteq G - L$ such that $T \cup F$ is a tour subgraph of $G$.
\end{definition}
We consider $L$  partial tour subgraphs defined by a given horizontal line $i$ and a given vertical line $j$ and refer to them as $L_{ij}$ partial tour subgraphs. Let $L_{ij}$ be the \textbf{induced} subgraph of $G$ consisting of vertices located in the rectangle $[1,j]\times[1,i]$ and the rectangle $[1,j-1]\times[i+1,h]$. More precisely, $L_{ij} = (V_{ij}, E_{ij})$ with $i \in [1, h]$ and $j \in [1, v]$ is the \textbf{induced} subgraph of $G$ defined by:
$$V_{ij} = \{v_{l,k} \in V \:|\: l \leq i, k \leq j\} \cup \{v_{l,k} \in V \:|\: l \geq i+1, k \leq j-1\}$$

Fig.~\ref{hanan} shows $L_{4,5}$ in a light gray area. $L_{4,5}$ contains the vertices in the rectangles $[1, 5]\times[1,4]$ and $[1,4]\times[5,7]$ since $h = 7$. Fig.~\ref{hanan} shows also an $L_{4,5}$ partial tour subgraph (black lines) since we can complete it (dashed lines) to obtain a complete tour subgraph of $G$. 

The right border of $L_{ij}$ is denoted $R_{ij}$ and defined as the subset of the $h$ rightmost vertices of $L_{ij}$ for each horizontal line. Fig.~\ref{hanan} represents the vertices of $R_{4,5}$ by squares inside the dark gray area as opposed to the light gray area representing $L_{4,5}$.

The rationale for the dynamic programming approach is the following: an optimal subtour consistent with $R_{4,5}$ consists of an optimal partial tour for the vertices to the left of $R_{4,5}$ combined to an optimal partial tour for the vertices to the right of $R_{4,5}$. The two restricted problems are made independent if enough information, namely the degree parity and connected components, is known about the vertices of $R_{4,5}$.

The algorithm will thus start with the initial state defined as the first partial subtour $L_{1,1}$. It extends the partial subtour $L_{ij}$ by adding vertical/horizontal edges between vertices iteratively. At the end, it has built the $L_{hv}$ partial tour subgraphs, which are all the possible tour subgraphs, so that a shortest can be identified as an optimal solution. We now describe the possible states and transitions between them.

\subsection{States.}
\label{statessec}
Two $L_{ij}$ partial tour subgraphs are considered equivalent if any completion of one of them is also a completion for the other.
The equivalent classes of $L_{ij}$ partial tour subgraphs can be characterized by some features of the vertices in $R_{ij}$. Let's go back to Fig.~\ref{hanan} and notice that the $L_{4,5}$ partial tour subgraph is made of two connected components. 
\begin{table}[t!]
\small
\begin{center}
$\begin{array}{|c|c|c|c|}
\hline
\textrm{no zero-degree} & \textrm{1 zero-degree} & \textrm{2 zero-degree} & \textrm{3 zero-degree} \\
\hline
\{(E, E, E), (1, 2, 3)\}  & \{(E, E, 0), (1, 2, -)\}  & \{(E, 0, 0), (1, -, -)\} & \{(0, 0, 0), (-, -, -)\}\\
\{(U, U, E), (1, 1, 2)\}& \{(0, E, E), (-, 1, 2)\} & \{(0, E, 0), (-, 1, -)\}  & \\
\{(E, E, E), (1, 1, 2)\} &\{(E, 0, E), (1, -, 2)\} &  \{(0, 0, E), (-, -, 1)\}  & \\
\{(E, U, U), (1, 2, 2)\} & \{(U, U, 0), (1, 1, -)\}  & & \\
\{(E, E, E), (1, 2, 2)\} & \{(E, E, 0), (1, 1, -)\} & & \\
\{(U, E, U), (1, 2, 1)\} &\{(0, U, U), (-, 1, 1)\}  & & \\
\{(E, E, E), (1, 2, 1)\}& \{(0, E, E), (-, 1, 1)\} & & \\
\{(U, U, E), (1, 1, 1)\} & \{(U, 0, U), (1, -, 1)\} & & \\
\{(E, U, U), (1, 1, 1)\} & \{(E, 0, E), (1, -, 1)\}  & & \\
\{(U, E, U), (1, 1, 1)\} & & & \\
\{(E, E, E), (1, 1, 1)\} & & & \\
\hline
\end{array}
$
\vspace{0.1cm}
\caption{States of $\Omega(3)$ (for RTSP) sorted by number of zero-degree vertices.\label{states}}
\label{tableStates}
\end{center}
\end{table}

So, to characterize the equivalent classes of $L_{ij}$ partial tour subgraphs, we need the degree parity of each vertex of $R_{ij}$ and distribution of vertices of $R_{ij}$ over distinct connected components. Two vertices of $R_{ij}$ can be connected with 0, an odd, or an even number of paths and the total number of incident paths determines the parity 0, U, or E of the vertex.  We use the same notation as \cite{rat83} to describe degree parities: even = E, odd = U (uneven) and zero = 0. Connected components are described by their indices or "$-$" for a zero degree. An equivalent class is related to a state of the dynamic program. In the following, such a state $\omega$ is denoted $\omega = \{(x_1, \ldots, x_h),(c_1, \ldots, c_h)\}$ where $x_i \in \{U,E,0\}$ and $c_i$ are respectively the parity label and the connected component of the $i$-th vertex of $\omega$. Moreover $C_j$ denotes the set of vertices belonging to connected component number $j$ so that $C_j = \{ i | c_i = j, \forall i = 1 \ldots h\}$. For example the equivalence class of the $L_{4,5}$ tour subgraph of Fig.~\ref{hanan} is described by the the following pair of vectors $\{(E, E, E, U, U, 0, 0), (1, 1, 1, 2, 2, -, -)\}$ and we have $C_1 = \{1,2,3\}$, $C_2 = \{4,5\}$. In this example, the vertices belong to two distinct connected components (of the corresponding class of tour subgraphs). In the first component (index 1), the three vertices are connected to an even number of paths and in the second component, the two vertices are connected to an odd number of paths.

We denote by $\Omega(h)$ the set of all possible states for $h$ vertices in the same $R_{ij}$, and $\Omega$ the set of all states of the dynamic program. Table~\ref{tableStates} lists all the possible states of $\Omega(3)$. We summarize a number of key observations about valid states that are needed to fully understand the algorithm and to establish its complexity.

Consider a connected component $j$ with a single vertex $i$ so that $C_j = \{i\}$. Such a vertex is referred as a vertex with \textbf{a path-reversal}.

\begin{figure}
   \begin{minipage}[c]{.48\linewidth}
	\centering
	\includegraphics[scale=0.9]{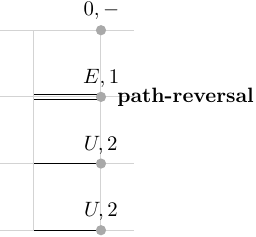}
	\caption{Example of a path-reversal: a vertex with two incoming edges and alone in its connected component.}
	\label{figLemma2}
   \end{minipage} \hfill
   \begin{minipage}[c]{.48\linewidth}
      	\centering
	\includegraphics[scale=0.9]{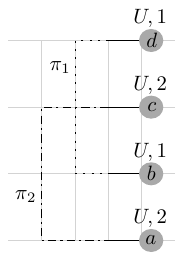}
	\caption{Example of an inconsistent labelling of the connected components. $\pi_1$ and $\pi_2$ must cross so that a,b,c and d must belong to the same connected component.}
	\label{figLemma4}
   \end{minipage}
\end{figure}

\begin{lemma}
A vertex with a path-reversal has degree 2.
\end{lemma}
\proof
The degree of $i$ is not zero since it belongs to a connected component. The only possible connection to vertex $i$ is from the vertex located to its left and it can have at most 2 edges. It has exactly 2 edges since the degree must be even, see Theorem \ref{th1}. $\hfill \square$
\endproof
An example of a path-reversal in a state is shown in Fig.~\ref{figLemma2}.

\begin{lemma}
\label{evenu}
A connected component of a state in $\Omega(h)$ has zero or an even number of vertices labeled with $U$.
\end{lemma}
\proof
By Lemma \ref{th1}, vertices of $L_{i,j} - R_{i,j}$ have an even degree. So in a state, only vertices of $R_{i,j}$ can have an odd degree. Since in a graph, the number of vertices of odd degree is even, a state has zero or an even number of vertices with an odd degree. $\hfill \square$

\endproof

\begin{lemma}\label{noncross}
The partition describing the connected components $\{C_1, \dots, C_k\}$ of a state is a non-crossing partition, \emph{i.e.}  if $a < b < c < d$ (ordered from bottom to top) and $a, c \in C_i$ and $b, d \in C_j$, then $i=j$.
\end{lemma}
\proof
Since vertices $a$ and $c$ belong to the same component $C_i$, there is a path $\pi_1$ from $a$ to $c$. Similarly,  there is another path $\pi_2$ from $b$ to $d$ (see Fig.~\ref{figLemma4}). Thus, $\pi_1$ and $\pi_2$ must cross, the intersection point is a vertex in both $\pi_1$ and $\pi_2$, thus $a, b, c$ and $d$ belong to the same connected component.  $\hfill \square$
\endproof

\subsection{Transitions.}
\label{sectrans}
\begin{figure}[t]
\begin{center}
\includegraphics[scale=1.2]{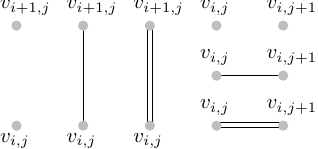}
\caption{Vertical and horizontal transitions}
\label{transitions}
\end{center}
\end{figure}

\begin{figure}
\begin{center}
\includegraphics[scale=0.92]{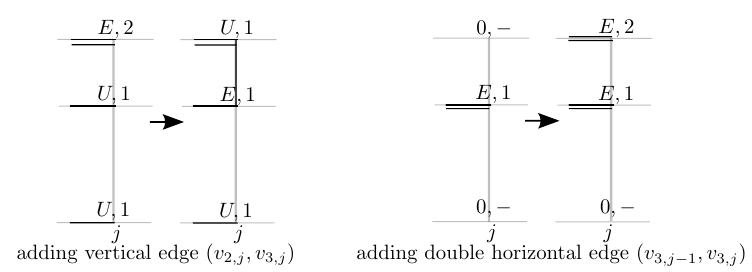}
\caption{Examples of two transitions: using a single vertical edge (left picture) and using a double horizontal edge (right picture).\label{extrans}}
\end{center}
\end{figure}

There are two types of transitions between states: vertical and horizontal transitions, corresponding to the addition of vertical or horizontal edges of $\Gamma(P)$ (see Fig.~\ref{transitions}).
In an optimal tour subgraph, two adjacent vertices in $G$ can be connected by one or two edges, or not connected.
The cost of a transition is the sum of the lengths of the edges. For instance (see Fig.~\ref{extrans}), the addition of a single vertical edge $(v_{2,j}, v_{3,j})$ to state $\{(U, U, E), (1, 1, 2)\}$ leads to the state $\{(U,E,U), (1,1,1)\}$. Similarly, the addition of a double horizontal edge $(v_{3,j-1}, v_{3,j})$ to state $\{(0, E, 0), (-, 1, -)\}$, leads to the state $\{(0, E, E), (-, 1, 2)\}$.

\subsection{Algorithm.}
\label{secalgo}
\begin{figure}[h!]
\begin{center}
 \includegraphics[scale=0.45]{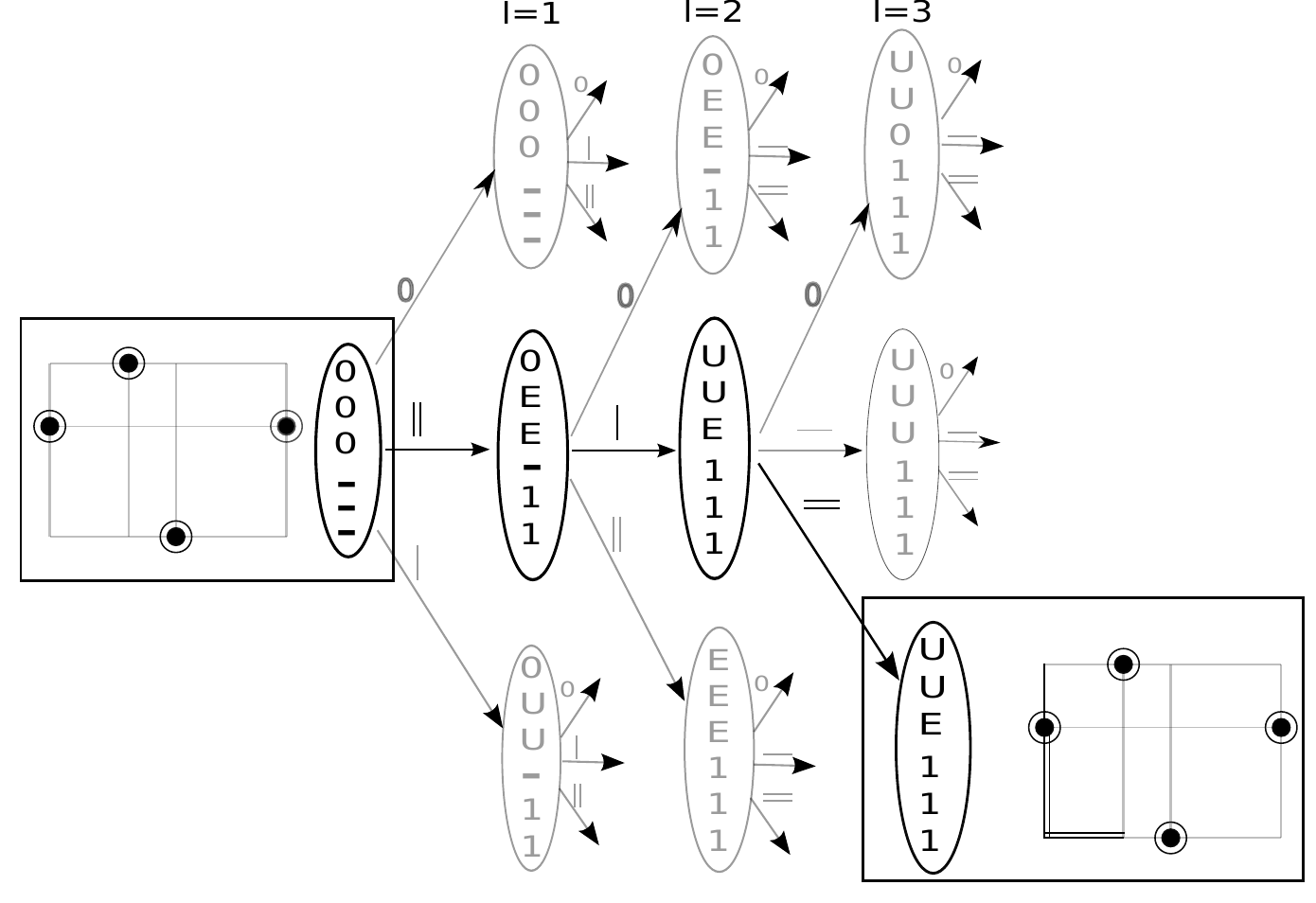}
\caption{Example of the graph underlying the dynamic programming algorithm. Each layer is identified with a value of $l$. Three transitions are possible from each state. The partial toursubgraph obtained by following the black path is shown on the bottom right corner.}
\label{gpdyn}
\end{center}
\end{figure}

Algorithm \ref{algogo} processes the edges of $\Gamma(P)$ from bottom to top and then from left to right (line \ref{line4}): the vertical edges $(v_{i, 1}, v_{i+1,1})$, then the horizontal edges $(v_{i,1}, v_{i,2})$, then the vertical edges $(v_{i, 2}, v_{i+1, 2})$ and so on. From any state (line \ref{line5}), three possible transitions are considered (line \ref{line6}): no edge, a single one or a double one. All the states obtained after adding $l$ transitions belong to the l-th layer and the dynamic programming algorithm can be seen as a shortest path algorithm in a layered graph (see Fig. \ref{gpdyn}). We denote by $T(\omega, l)$ the value of the shortest path to reach state $\omega$ located on layer $l$. 
\begin{algorithm}[h!]
\small
\begin{algorithmic}[1]
\STATE $\omega_0 \leftarrow \{(0,\ldots,0), (-,\ldots,-)\}$; $T(w_0, 0) = 0$; $Layer_0  \leftarrow \{w_0\}$ \label{line1}
\STATE $l \leftarrow 0$ \label{line2}
\FOR{each edges of $e \in \Gamma(P)$ from bottom to top and from left to right} \label{line3}
		\STATE $Layer_{l+1} \leftarrow \emptyset$ \label{line4}
		\FOR{each state $\omega \in Layer_l$} \label{line5}			
			\FOR{each possible transitions (zero, one or two edges) $tr$ for $e$}  \label{line6}
				\STATE $\omega' \leftarrow \omega + tr$  \label{line7}
		\IF{check($\omega', l+1$)} \label{lcheck}
					\IF{$\omega' \in Layer_{l+1}$} \label{line9}
						\IF{$T(\omega', l+1) > T(\omega, l) + length(tr)$} \label{line10}
							\STATE $T(\omega', l+1)\leftarrow T(\omega, l) + length(tr)$ \label{line11}
						\ENDIF
					\ELSE
						\STATE $T(\omega', l+1)\leftarrow T(\omega, l) + length(tr)$
						\STATE $Layer_{l+1} \leftarrow Layer_{l+1} \cup \{\omega'\}$  \label{line14}
					\ENDIF 
				\ENDIF
			\ENDFOR
		\ENDFOR
		\STATE $l \leftarrow l+1$
\ENDFOR
\STATE $w_{opt} \leftarrow \argmin_{\omega \in L_{hv}} T(\omega, l^*)$ where $l^*$ is the last layer.
\RETURN $w_{opt}$
\caption{Dynamic Programming algorithm for the Rectilinear TSP}
\label{algogo}
\end{algorithmic}
\end{algorithm}

Lines \ref{line7}-\ref{line14} updates the possible states of the next layer (l+1) by extending the considered state $\omega$ of layer $l$ with the considered transition $tr$. The new state $\omega'$ might not be a valid tour subgraph and it is checked line \ref{lcheck}. Lines \ref{line9}-\ref{line14} are the traditional update of the shortest path values. Typically line \ref{line9} checks whether an existing partial tour subgraph is already known to reach $\omega'$ on layer $l+1$. If yes, it is checked line \ref{line10}  whether a shorter one has been found and $T(\omega', l+1)$ is updated accordingly. 
An illustrative execution of the algorithm is shown Fig.~\ref{gpdyn}, where a particular path is outlined demonstrating the relation between states and partial tours. The algorithm is a shortest path algorithm in the graph of Fig.~\ref{gpdyn} where there are $O(hv)$ layers (the maximum number of possible transitions), at most $|\Omega(h)|$ nodes in each layer and at most three outgoing arcs from any node.

Let's give more details about line \ref{lcheck}  and infeasible or suboptimal states. The following conditions come directly from Theorem \ref{th1} and must be satisfied by the states to ensure the algorithm computes a valid tour subgraph:
\begin{enumerate}
\item A non-zero degree vertex has an even number of incident edges in any state.
\item A vertex belonging to $P$ has a positive degree in any state.
\item A state located on the last layer must have a single connected component since the tour subgraph must be connected. 
\end{enumerate}

Conditions 1 and 2 can be checked after the last step involving an incident edge of the vertex and condition 3 after all edges of $\Gamma(P)$ have been processed. These conditions are evaluated when considering a state $\omega$ for layer $l$ (see the call to $check(\omega,l)$ line \ref{lcheck} of Algorithm \ref{algogo}) to filter invalid states. Additionally we know that some states cannot belong to an optimal tour sub-graph. The following simple filtering rule is applied to speed up the algorithm and rule out some sub-optimal states:
\begin{itemize}
\item A vertex $v_{i, j}$ that is not in $P$ can not be solely connected to two parallel edges (this would create a useless turn back and forth in $v_{i, j}$).
\end{itemize}

\subsection{Complexity analysis.}
\label{complexity}
All states of the graph underlying the dynamic program have an out degree of at most three. The time-complexity of the algorithm solely depends on the number of states. Since we have $O(hv)$ layers (exactly $(h-1)v+(v-1)h = 2hv-h-v$), we focus on counting the maximum number of states of a layer. Notice that Figure \ref{gpdyn} implies a total of $3^{O(hv)}$ states but some of them are identical and we refine the counting in the present section.
The number of possible states made of $h$ possible vertices is denoted by $|\Omega(h)|$. As an example, Table~\ref{tableStates} which enumerates the states belonging to $\Omega(3)$ shows that $|\Omega(3)| = 24$. 
To compute $|\Omega(h)|$, we proceed as follows: 

\begin{enumerate}
\item Firstly, we compute $|\Omega_{pos}(h)|$, the number of states with \textbf{only positive degree vertices} (\emph{i.e.} without vertices of zero-degree). We thus have $\Omega_{pos}(h) \subset \Omega(h)$. We will show that $|\Omega_{pos}(h)|$ is equal to the super Catalan number by using a particular interpretation of the super Catalan number referred to as $f^4$ in \cite{fan}: \\

$f^4_n$: {\it Number of ways of connecting $n$ points in the plane lying on a horizontal line by noncrossing edges above the line such that if two edges share an endpoint $p$, then $p$ is a left endpoint of both edges. Then color each edge by black or white.} \\

We denote by $\Omega_{f^4}(h)$ the set of configurations described above and establish a bijection between $\Omega_{pos}(h)$ and $\Omega_{f^4}(h)$ (Lemma \ref{lembij}). Since $f^4_n$ is known to be equal to the super Catalan number $S_{n}$ (see \cite{fan}), we have $|\Omega_{pos}(h)| = S_{h}$ (Lemma \ref{theo1}).\\

\item Secondly, we consider the zero-degree vertices to relate $|\Omega(h)|$ to the super Catalan numbers (Lemma \ref{lemhs}). We can then prove (Theorem \ref{theoBnd}) that $|\Omega(h)| = O\left(\frac{(4+\sqrt{8})^{h+1}}{\sqrt{(h+1)^{3}}}\right)$. 
\end{enumerate}

\begin{figure}
\begin{center}
\includegraphics[scale=0.9]{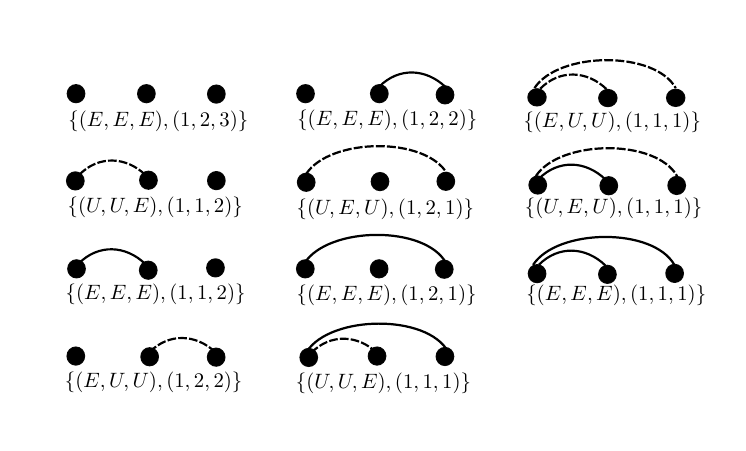}
\caption{Example of bijection between $\Omega_{pos}(3)$ and $\Omega_{f^4}(3)$.}
\label{bijection}
\end{center}
\end{figure}

We now start by establishing a bijection between the number of states with vertices of positive degree and the configurations of $f^4$. An example of such a bijection for $h=3$ is given in Fig.~\ref{bijection}. The 11 states of $\Omega_{pos}(3)$  are taken from the first column of table \ref{states} and are matched to the 11 configurations of $\Omega_{f^4}(3)$. The key ideas of the bijection between two states $\omega \in \Omega_{pos}$ and $\omega' \in \Omega_{f^4}$ are the following: a left endpoint and all points connected to it in $\omega'$ encode a connected component of  $\omega$; the color black/white of an edge $(i,j)$ of $\omega'$ relates to the parity of the number paths incident to $j$ in $\omega$. 

\begin{lemma}
\label{lembij}
There is a bijection between $\Omega_{pos}(h)$ and $\Omega_{f^4}(h)$.
\end{lemma}

\proof 
Recall that a state $\omega \in \Omega_{pos}$ is denoted $\omega = \{(x_1, \ldots, x_h),(c_1, \ldots, c_h)\}$. A state $\omega' \in \Omega_{f^4}$ is described by $h$ consecutive points $p_1, \ldots, p_h$ on a line and a set of white/black edges. In the following, $p_i$ is the point associated to the $i$-th  vertex of $\omega$.

Let's define an application $F$ from a state $\omega \in \Omega_{pos}$ to a state $\omega' \in \Omega_{f^4}$ as follows. Firstly, we consider each connected component $C$ of $\omega$ consisting of more than one vertex \emph{i.e.} of vertices $v_1, v_2, \dots, v_k$ ($c_{v_1} = c_{v_2} = \ldots = c_{v_k}$). For each vertex $v_i \in C$ with $i \neq 1$, if $x_{v_i}$ is a $U$ (resp. $E$) vertex, we add a white (resp. black) edge between the points $p_{v_1}$ and $p_{v_i}$. Secondly, a path-reversal vertex $v_i$ is matched to a zero-degree point $p_{v_i}$. Edges of $w'$ can only share their left endpoint and since the connected components of $w$ are non-crossing partitions (Lemma \ref{noncross}), the edges of $w'$ are non-crossing. Thus $\omega' \in \Omega_{f^4}$.

Let's now show that $F$ is injective. Consider two states $\omega_1$ and $\omega_2$  of $\Omega_{pos}$ and suppose that $F(\omega_1) = F(\omega_2)$. 
\begin{itemize}
\item $\omega_1$ and $\omega_2$ have the same connected components since $w$ and $F(w)$ have the same connected components by construction.
\item  The labels U/E of a vertex $v_i \neq v_1$ inside a connected component $v_1, v_2, \dots, v_k$ are in bijection with the color white/black of the edge $(p_{v_1},p_{v_i})$. Thus all such vertices in $\omega_1$ and $\omega_2$ have the same parity labels.
\item Since the number of $U$ labels must be even (Lemma \ref{evenu}), the label $x_{v_1}$ in each connected component $v_1, v_2, \dots, v_k$ is determined by the labels of $x_{v_2}, \ldots, x_{v_k}$.
\end{itemize}
Therefore $\omega_1 = \omega_2$ and $F$ is injective.

Let's show that $F$ is surjective. Consider $\omega' \in \Omega_{f^4}$ and let's show that there exists  $\omega \in \Omega_{pos}$ such that $F(\omega) = \omega'$. We build $\omega$ by defining $F^{-1}$ as follow.  Firstly, we consider each connected component of  $\omega'$ consisting of more than one vertex \emph{i.e} of vertices $v'_1, \ldots, v'_k$. For each point $p_{v'_i}$ with $i \neq 1$, if the edge connected to $p_{v'_i}$ is white (resp. black), we set $x_{v'_i}$ to $U$ (resp. $E$). The label $x_{v'_1}$ is set to $E$ if the number of white edges connected to $p_{v'_1}$ is even, or $U$ otherwise. Secondly, the label of a zero degree point is to $E$. Finally, we index connected components in increasing order.  
We now check that $\omega \in \Omega_{pos}$. First, the connected components of $\omega$ are non-crossing since the edges of $\omega'$ do not cross. Then,  we check that the number of $U$ labels in a connected component of $\omega$ is even (Lemma \ref{evenu}). Indeed, in a connected component $v_1, \ldots, v_k$, the number of white edges is in bijection with the number of $U$ labels of $x_{v_2}, \ldots, x_{v_k}$ and the label $x_{v_1}$ is chosen so that the total number of labels $U$ is even.

$F$ is therefore a bijective application. $\hfill \square$
\endproof

\begin{lemma}
\label{theo1}
$|\Omega_{pos}(h)| = S_{h}$, where $S_n$ is the super Catalan number (see OEIS A001003).
\end{lemma}
\proof
$|\Omega_{pos}(h)|= f^4_h$ due to Lemma \ref{lembij} and $f^4_h$ is known (see \cite{fan}) to be equal to the super Catalan number $S_{h}$. $\hfill \square$
\endproof

We now include the zero-degree vertices in the counting to obtain $|\Omega(h)|$.
\begin{lemma}
\label{lemhs}
$|\Omega(h)|= \sum^{h}_{k=0} {h\choose k} S_{k}$
\end{lemma}
\proof
When $(h-k)$ vertices out of $h$ have a zero-degree, there are $|\Omega_{pos}(k)|$ ways to connect the $k$ remaining vertices. This is because vertices with zero-degree are completely independent from the other vertices. The number of states with exactly $(h-k)$ zero-degree vertices is thus ${h \choose h-k} |\Omega_{pos}(k)|$. So $|\Omega(h)| = \sum^{h}_{k=0} {h\choose h-k} |\Omega_{pos}(k)|= \sum^{h}_{k=0} {h\choose k} |\Omega_{pos}(k)|$. By Lemma \ref{theo1} we end up with $|\Omega(h)|= \sum^{h}_{k=0} {h\choose k} S_{k}$.  $\hfill \square$

\endproof

The problem boils down to computing a bound on the sequence defined in Lemma \ref{lemhs} and based on the super Catalan numbers. 
\begin{theorem}
\label{theoBnd}
$|\Omega(h)| = O\left(\frac{(4+\sqrt{8})^{h+1}}{\sqrt{(h+1)^{3}}}\right)$
\end{theorem}
\proof
First we show that $|\Omega(h)| = T_{h+1}$ where $T_n$ is a specific number defined in OEIS A118376~\footnote{The interpretation of $T_n$, which is irrelevant to the proof, is given as \emph{the number of all trees of weight n, where nodes have positive integer weights and the sum of the weights of the children of a node is equal to the weight of the node}.}  (see \cite{oeis}).

Let $A(x)$ and $B(x)$ be the generating functions of respectively $T_n$ and $S_n$ (the super catalan numbers) \emph{i.e.} $A(x) = \sum_{n\ge 0} T_nx^n$ and $B(x) = \sum_{n\ge 0} S_nx^n$. Closed forms are known for both functions so that $A(x) = \frac{1-\sqrt{8x^2-8x+1}}{4-4x}$ and $B(x) = \frac{1+x-\sqrt{1-6x+x^2}}{4x}$  (see \cite{oeis}). As a result, we can express $A(x)$ as a composition of $B(x)$ and $\frac{x}{1-x}$ as follow $A(x) = \frac{x}{1-x}B(\frac{x}{1-x})$.

We use a result from \cite{kru14} to compute compositions of generating functions and adapt the proof of {\bf Theorem~8} of \cite{kru14}:
$$A(x) = \frac{x}{1-x} B\left(\frac{x}{1-x}\right) = \frac{x}{1-x} \sum_{k\ge 0} S_k \left(\frac{x}{1-x}\right)^k = \sum_{k\ge 0} S_k \left(\frac{x}{1-x}\right)^{k+1}$$

Replacing $\left(\frac{x}{1-x}\right)^{k+1}$ by $\sum_{n\ge k+1} {n-1 \choose n-k-1}x^n$ (see \cite{kru14}) we obtain:

\begin{align*}
   A(x)  = {0 \choose 0} S_0 x +  {1 \choose 0} S_0 x^2 +  {2 \choose 0} S_0 x^3 +  \dots + & {n-1 \choose 0} S_0 x^n +  \dots \\
  + {1 \choose 1} S_1 x^2 +  {2 \choose 1} S_1 x^3 +  \dots + & {n-1 \choose 1} S_1 x^n +  \dots \\
  + {2 \choose 2} S_2 x^3 +  \dots + &  {n-1 \choose 2} S_2 x^n +  \dots \\
 & \dots \\
+ & {n-1 \choose n-1} S_{n-1} x^n + \dots
\end{align*}

Summing the coefficients of $x^n$ for $n>0$, we get $A(x) = \sum_{n\ge 0} T_n x^n$ where

$$T_{n}= \sum^{n}_{k=1} {n-1\choose k-1} S_{k-1} = \sum^{n-1}_{k'=0} {n-1\choose k'} S_{k'}$$

By taking $n = h+1$, we can state $T_{h+1} = \sum^{h}_{k'=0} {h\choose k'} S_{k'}$ so we have, from Lemma \ref{lemhs}, $|\Omega(h)| = T_{h+1}$. Moreover, we can use the singularity method given in \cite{fla09} (Proposition IV.1 and Theorem V1.1) to show that $T_{h+1}$ is in $O\left(\frac{(4+\sqrt{8})^{h+1}}{\sqrt{(h+1)^{3}}}\right)$. $\hfill \square$
\endproof

\begin{table}[h!]
\small
\begin{center}
\begin{tabular}{|c|c|c|c|c|c|c|c|c|c|c|}
\hline
$h$     		   & 1 & 2 & 3 & 4 & 5 & 6 & 7 & 8 & 9 & 10\\ \hline
$|\Omega_{pos}(h)|= S_{h}$ & 1 & 3 & 11 & 45 & 197 & 903 & 4279 & 20793 & 103049 & 518859 \\ \hline
$|\Omega(h)|= T_{h+1}$ & 2 & 6 & 24 & 112 & 568 & 3032 & 16768 & 95200 & 551616 & 3248704 \\
\hline
\end{tabular}
\vspace{0.2cm}
\caption{The exact value of the numbers $|\Omega(h)|$ and $|\Omega_{pos}(h)|$ for $h$ between 1 and 10.\label{tGH}}
\label{orders}
\end{center}
\end{table}
The number of states $|\Omega(h)|$ of the dynamic program is thus in $O\left(\frac{(4+\sqrt{8})^{h+1}}{\sqrt{(h+1)^{3}}}\right)$. Table~\ref{orders} shows the order of magnitude of the numbers involved. Since there are at most $O(hv)$ edges to consider, the overall time complexity of algorithm \ref{algogo} is in $O\left(\frac{(4+\sqrt{8})^{h+1}}{\sqrt{(h+1)^{3}}}hv\right)$ assuming that we can check that a state belongs to a layer in constant time (line 9 of Algorithm \ref{algogo}). For sake of simplicity, we highlight $n$ and $h$ only (since $v \leq n$) and simplify the complexity to $O\left(hn7^h\right)$.

\section{Fixed-parameter algorithm for Rectilinear Steiner tree.}
\label{algosteiner}
\begin{figure}[b!]
\begin{center}
\includegraphics[scale=0.8]{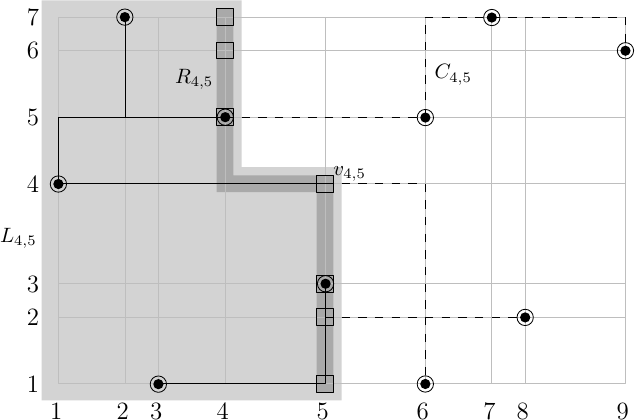}
\caption{Hanan grid $\Gamma(P)$ with $L_{4,5}$ in the light gray area and $R_{4,5}$ in the dark gray area. The black edges represent an $L_{4,5}$ partial tree and the dashed edges represent one possible completion.}
\label{rectSteiner}
\end{center}
\end{figure}

We now apply the exact same methodology to the Steiner tree problem. We briefly describe how each step is modified to handle the Steiner tree case. Notice that the previous methodology was described in details for the more complex case of rectilinear TSP and that it is now merely simplified. For sake of simplicity the notations are kept identical. We now define the undirected graph $G = (V,E)$ by associating a vertex to each intersection of $\Gamma(P )$ and a single edge for each segment of $\Gamma(P)$, with length equal to the $l_1$ distance between the intersections. The L partial tour subgraph become L partial trees.

\begin{definition}\label{d1}
For any subgraph $L \subseteq G$; a tree $T \subseteq L$ is a $L$ \textbf{partial tree} if there exists a tree $F \subseteq G - L$ such that $T \cup F$ is a Steiner tree of $G$.
\end{definition}

Figure \ref{rectSteiner} demonstrates an $L_{ij}$ partial tree and its $R_{ij}$ rightmost frontier. An $L_{45}$ partial tree is shown and a possible completion to a complete Steiner tree. 

\subsection{States and Transitions.}
\label{statessec_RST}
\paragraph{States} A state $\omega$ is denoted $\omega = (c_1, \ldots, c_h)$ where $c_i$ is the connected component of the $i$-th vertex of $\omega$. Connected components are described by their indices or "$-$" for a zero degree. Notice that the parity of the degree does not need to be stored anymore. Regarding the degree information, we now only need to know whether the degree is null or not. This information amounts at checking whether $c_i \neq $"$-$". Table~\ref{statesSteiner} gives the possible states for $h=3$. For the exact same reason given in section \ref{statessec}, the partition describing the connected components $\{C_1,\ldots,C_k\}$ of a state is a non-crossing partition.
Typically, the state representing the equivalent class of the $L_{4,5}$ partial tree of Fig.~\ref{rectSteiner} is described by the following vector $(1, 1, 1, 2, 2, -, -)$. 
\begin{table}[t!]
\small
\begin{center}
$\begin{array}{|c|c|c|c|}
\hline
\textrm{0 zero-degree} & \textrm{1 zero-degree} & \textrm{2 zero-degree} & \textrm{3 zero-degree} \\
\hline
(1, 1, 1) & (1, 1, -) & (1, -, -) & (-, -, -) \\
(1, 1, 2) & (-, 1, 1) & (-, 1, -) & \\
(1, 2, 1) & (1, -, 1) & (-, -, 1) & \\
(1, 2, 2) & (1, 2, -) & 	    & \\
(1, 2, 3) & (1, -, 2) & & \\
	     &  (-, 1, 2)  & & \\
\hline
\end{array}
$
\vspace{0.1cm}
\caption{States of $\Omega(3)$ for the Steiner Tree.\label{statesSteiner}}
\label{tableStates_tree}
\end{center}
\end{table}

\paragraph{Transitions} There are two types of transitions: vertical and horizontal. However, there is now only two possible configurations for connecting two adjacent vertices of $G$ in an optimal Steiner tree: zero or one edge.

\subsection{Algorithm.}
 Algorithm \ref{algoSteiner} gives the pseudo-code where the changes compared to Algorithm \ref{algogo} are highlighted by rectangular boxes.
The algorithm is only modified lines \ref{line1}, \ref{line6} and \ref{lcheck}. The modification lines \ref{line1} and \ref{line6} simply account for the new definition of the states and the restriction of the transitions to two cases (rather than three for the TSP).
\begin{algorithm}[h!]
\small
\begin{algorithmic}[1]
\STATE  \framebox{$\omega_0 \leftarrow (-,\ldots,-)$;} $T(w_0, 0) = 0$; $Layer_0  \leftarrow \{w_0\}$ \label{line1}
\STATE $l \leftarrow 0$
\FOR{each edges of $e \in \Gamma(P)$ from bottom to top and left to right} \label{line4}
		\STATE $Layer_{l+1}  \leftarrow \emptyset$
		\FOR{each state $\omega \in Layer_l$} \label{line5}			
			\FOR{each possible transitions \framebox{(zero or one edge)} $tr$ for $e$}  \label{line6}
				\STATE $\omega' \leftarrow \omega + tr$  \label{line7}
		\IF{ \framebox{check($\omega', l+1$)}} \label{lcheck}
					\IF{$\omega' \in Layer_{l+1}$}
						\IF{$T(\omega', l+1) > T(\omega, l) + length(tr)$} \label{line10}
							\STATE $T(\omega', l+1)\leftarrow T(\omega, l) + length(tr)$
						\ENDIF
					\ELSE
						\STATE $T(\omega', l+1)\leftarrow T(\omega, l) + length(tr)$
						\STATE $Layer_{l+1} \leftarrow Layer_{l+1} \cup \{\omega'\}$  \label{line14}
					\ENDIF
				\ENDIF
			\ENDFOR
		\ENDFOR
		\STATE $l \leftarrow l+1$
\ENDFOR
\STATE $w_{opt} \leftarrow \argmin_{\omega \in L_{hv}} T(\omega, l^*)$ where $l^*$ is the last layer
\RETURN $w_{opt}$
\caption{Dynamic programming algorithm for Steiner Tree}
\label{algoSteiner}
\end{algorithmic}
\end{algorithm}
Removing infeasible states line \ref{lcheck} amounts at checking that:
\begin{enumerate}
\item A vertex of $P$ have a positive degree
\item The partial tree is connected (all vertices on the last layer belong to the same connected component)
\end{enumerate}

The following rules are also applied to filter sub-optimal or symmetrical states. We basically restrict the algorithm to identify trees that satisfy the following conditions:
\begin{enumerate}
\item A vertex $v_{i, j}$ that is not in $P$ can not be connected to a single edge (it would create a useless pendant vertex).
\item Two vertices already in the same connected component can not be directly connected (it would create a cycle which is sub-optimal).
\item Two horizontally (resp. vertically) adjacent vertices $v_{i, j}$ anf $v_{i, j+1}$ (resp $v_{i, j}$ and $v_{i+1, j}$) that both belongs to $P$ are connected by the direct corresponding edge $(v_{i, j}, v_{i, j+1})$ (resp. $(v_{i, j}, v_{i+1, j})$) \cite{aho1977}.   
\item Any horizontal/vertical line (sequence of consecutive horizontal/vertical edges) contains at least one point of $P$ \cite{Brazil2000}.
\end{enumerate}

Conditions (4) is global and requires, for efficient checking, to store whether a vertex in the state is connected to a point of $P$. 
 
\subsection{Complexity.}
\label{complexitycatalan}
Any state in the graph underlying the dynamic program has now a degree of at most two and there are $O(hv)$ layers in the graph. We therefore establish the complexity of the algorithm by counting the number of possible states $|\Omega(h)|$ of a single layer. As an example, Table~\ref{tableStates_tree} enumerates the states belonging to $\Omega(3)$ so $|\Omega(3)| = 15$. $|\Omega(h)|$ is computed as follows:

\begin{enumerate}
\item Firstly, we show that $|\Omega_{pos}(h)|$ is equal to the Catalan number referred to as $h^5$ in \cite{sta}: \\

$h^5_n$: {\it Ways of connecting $n$ points in the plane lying on a horizontal line by noncrossing arcs above the line such that if two arcs share an endpoint $p$, then $p$ is a left endpoint of both the arcs.} \\

Since $\Omega_{pos}(h)$ is in bijection with $\Omega_{h^5}(h)$ and $h^5_n$ is known to be equal to the Catalan number $C_{n}$ (see \cite{sta}), we have $|\Omega_{pos}(h)|= C_{h}$.\\

\item Secondly, we prove that $|\Omega(h)|$  is a known sequence related to the Catalan numbers (see OEIS A007317 \cite{oeis}).

\end{enumerate}

\begin{figure}
\begin{center}
\includegraphics[scale=0.9]{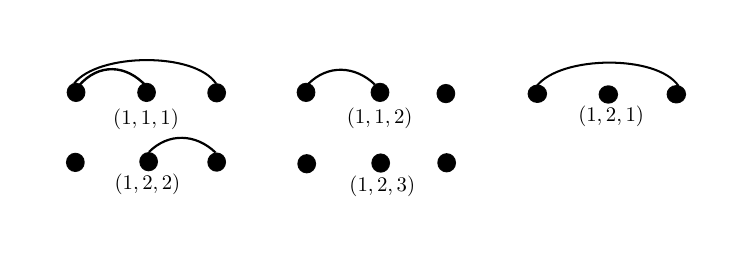}
\caption{Example of bijection between $\Omega_{pos}(3)$ and $\Omega_{h^5}(3)$. Note that at most one single connected component appears for $h=3$ when no path-reversal vertex is allowed.}
\label{bijection_Steiner}
\end{center}
\end{figure}

We now start by establishing a bijection between the number of states with a positive degree and the configurations of $h^5$. An example of such a bijection for $h=3$ is given Fig.~\ref{bijection_Steiner}. The 5 states of $\Omega_{pos}(3)$ are taken from the first column of Table~\ref{tableStates_tree} and are matched to the 5 configurations of $\Omega_{h^5}(3)$.

\begin{lemma}
\label{lembij_steiner}
There is a bijection between $\Omega_{pos}(h)$ and $\Omega_{h^5}(h)$.
\end{lemma}

\proof
Recall that a state $\omega \in \Omega_{pos}$ is denoted $\omega = (c_1, \ldots, c_h)$ (see section \ref{statessec_RST}). A state $\omega' \in \Omega_{h^5}$ is described by $h$ consecutive points $p_1, \ldots, p_h$ on a line and a set of edges. In the following, $p_i$ is the point associated to the $i$-th  vertex of $\omega$.

Let's define an application $F$ from a state $\omega \in \Omega_{pos}$ to a state $\omega' \in \Omega_{h^5}$ as follows. 
We consider each connected component $C$ of $\omega$ consisting of vertices $v_1, v_2, \dots v_k$ ($c_{v_1} = c_{v_2} = \ldots = c_{v_k}$). For each vertex $v_i \in C$ with $i \neq 1$, we add an edge between the points $p_{v_1}$ and $p_{v_i}$. We can check that $\omega'$ is a valid configuration of $\Omega_{h^5}$ (non crossing edges sharing only their left endpoint).
Moreover, there is a one to one direct correspondence between the connected components of $\omega$ indexed in increasing order and the connected components of $\omega'$. The application $F$ is therefore a bijection. $\hfill \square$
\endproof

\begin{lemma}
\label{theo1Tree}
$|\Omega_{pos}(h)|= C_{h}$, where $C_n$ is the Catalan number (see OEIS A000108).
\end{lemma}
\proof
$|\Omega_{pos}(h)|= h^5_h$ due to Lemma \ref{lembij_steiner} and $h^5_h$ is known (see \cite{sta}) to be equal to the Catalan number $C_{h}$. $\hfill \square$
\endproof

\begin{lemma}
\label{lemhsTree}
$|\Omega(h)|= \sum^{h}_{k=0} {h\choose k} C_{k}$
\end{lemma}
\proof
Identical to proof of Lemma \ref{lemhs}. $\hfill \square$

\endproof

\begin{theorem}
\label{theoBndTree}
$|\Omega(h)|= O\left(\frac{5^{h}}{\sqrt{h^{3}}}\right)$
\end{theorem}
\proof
The formula $\sum^{h}_{k=0} {h\choose k} C_{k}$ of Lemma \ref{lemhsTree} is known as the integer sequence A007317 (see \cite{oeis}). Since the generating function of this sequence is $\frac{3}{2}-\frac{1}{2}\sqrt{\frac{1-5x}{1-x}}$, we can use the singularity method described in \cite{fla09} (Proposition IV.1 and Theorem VI.1) to show that $|\Omega(h)| = O\left(\frac{5^{h}}{\sqrt{ h^{3}}}\right)$.
 $\hfill \square$
\endproof

There are $O(hv)$ layers and the number of states in a given layer is in $O(5^h)$ so the overall time-complexity can be expressed as $O(nh5^h)$ for sake of simplicity.

\section{Comparison and relationship to the rank based technique}
\label{rankbased}

The planar grid-graph used in this paper has a pathwidth and treewidth of $h$
and the rank based approach recently proposed by Bodlaender et al. \cite{Bodlaender2015} can be directly applied. It is indeed intended for graph problems with a bounded treewidth/pathwidth and a global connectivity property such as the Hamiltonian Cycle, Steiner tree or TSP. We focus on the results obtained via a path decomposition which are stronger for the present paper. A path decomposition of our grid-graph is made of bags (or separators) of at most $h+1$ vertices and the dynamic programming approach must encode some information about the degree as well as the connected components to which the vertices of a bag belong. Encoding the connected components involve partitioning the vertices of a bag (one partition refers to one component) which lead to consider a $h^h$ number of partitions. In a breakthrough result, the authors of \cite{Bodlaender2015} show that there exists a representative subset of these partitions of size $2^h$. Moreover, given a set $\mathcal{A}$ of partitions, it is possible to compute this representative subset $\mathcal{A^{'}} \subseteq \mathcal{A}$ (with $|\mathcal{A^{'}}| \leq 2^{h}$) in time $|\mathcal{A}|2^{(w-1)h}h^{O(1)}$ where $w$ is the matrix multiplication exponent (see Theorem 3.7 of \cite{Bodlaender2015}). Regarding parameter $w$, it refers to the complexity $O(n^w)$ for multiplying two $n$ by $n$ matrices. In brief, the best known upper bound for $w$ is $w=2.3727$ \cite{Williams2012} and $w=2$ remains the best known lower bound so far. The number of partitions generated in the course of the algorithm over a nice path decomposition at most doubles from one bag to the next (when an edge is introduced). Thus the algorithm is always applied on a set of at most $2^{h+1}$ partitions and the reduction is performed in time $2^{wh}h^{O(1)}$. 
We now briefly review the runtimes obtained with this technique for the two problems considered here. Table \ref{table_complexity} gives a summary of the comparison to the best knowledge of the authors.
\paragraph{Steiner Tree} Let us consider a bag of the path decomposition and count the states sharing a given set of $k$ positive degree vertices (there are at most $h+1\choose k$ of such sets in the bag). In each set, the number of states relates to the number of partitions of $k$ elements and is bounded by $k^k$ in general. It is here the number of non-crossing partitions counted by the Catalan numbers as explained in section \ref{complexitycatalan} and is thus bounded\footnote{This bound can be proved by using Stirling's approximation of $n!$ applied to the explicit formula for the Catalan numbers $C_n = \frac{1}{n+1}{2n \choose n}$} by $4^k$. Alternatively,  applying the rank-based reduction algorithm would reduce the number of such partitions to $2^k$ and thus reduce the space needed. But, as explained above, the reduction algorithm runs in $2^{wk}k^{O(1)}$. It follows that the time complexity to process all states of a bag is bounded by $\sum^{h+1}_{k=1} {h+1\choose k} 2^{wk}h^{O(1)}= (1+2^w)^{h+1}h^{O(1)}$. Overall, considering that the number of bags is linearly dependent of $n$, \cite{Bodlaender2015} reports a $n(1+2^{w})^{h}h^{O(1)}$ runtime. By assuming the best case of $w=2$, the complexity matches the $O(nh5^{h})$ proposed in this paper. Our approach thus improves over the rank-based technique for the current best known matrix multiplication algorithm with $w = 2.3727$ but takes advantage of the grid structure in addition to bounded pathwidth.

\paragraph{Steiner Traveling Salesman}
We apply the same reasoning to the case of the Steiner TSP. Consider all the states with a given triple of $k_0$ vertices of zero degree, $k_1$ vertices of odd degree and $k_2$ vertices of even degree. The number of such states relates to the number of partitions of $k_1+k_2$ elements. 
The rank based approach reduces the number of such partitions to $2^{k_1+k_2}$ and the complexity is bounded by (using the multinomial theorem):
$\sum\limits_{k_0 + k_1 + k_2 = h+1}  {h+1 \choose k_0, k_1, k_2} 1^{k_0} 1^{k_1}2^{(k_1+k_2)w} h^{O(1)} = (1 + 2^{w+1})^{h+1} h^{O(1)}$. The runtime is thus in $n(1 + 2^{w+1})^{h} h^{O(1)}$
and our approach improves over the rank-based technique even when assuming the best case of $w = 2$.

The Steiner TSP is not addressed in \cite{Bodlaender2015} but a $n(2+2^{w/2})^{h}h^{O(1)}$ runtime for TSP is reported. When considering the TSP, the partitions boil down to perfect matchings of the degree one vertices. It is known from \cite{Cygan2013} that an improved reduction algorithm can be applied in this specific case and the number of representative partitions is only $2^{h/2}$. However, the partitions considered for Steiner TSP are not strictly speaking matchings since they encompass the even degree vertices even though the odd degree vertices are paired. Thus the aforementionned result of \cite{Cygan2013} does not seem applicable. \\

\begin{table}[h!]
\begin{center}
\begin{tabular}{|c|c|c|c|}
\hline
Assumptions 	    & pathwidth $h$ 		 & pathwidth $h$ & grid graph of pathwith $h$\\
		 	    & ($2 \leq w < 2.3727$) &  ($w = 2$)        &	\\
\hline
Steiner Tree	    & $n(1+2^{w})^{h}h^{O(1)}$   & $n5^{h}h^{O(1)}$  & $O(nh5^{h})$  \\
Steiner TSP	    & $n(1+2^{w+1})^{h}h^{O(1)}$  & $n9^{h}h^{O(1)}$  &  $O(nh7^{h})$ \\
          TSP	    &  $n(2+2^{w/2})^{h}h^{O(1)}$  & $n4^{h}h^{O(1)}$  &  - \\

\hline
\end{tabular}

\vspace{0.1cm}
\caption{Comparison with the rank based approach.}
\label{table_complexity}
\end{center}
\end{table}

\section{Experimental results.}
\label{expe}
We performed simple experiments that serve as a proof of concept and show the scalability of the proposed algorithms. In particular, we show that the rectilinear TSP can be solved exactly up to $h=8$ horizontal lines in practice demonstrating that this algorithm could be used for real-life picking problems in warehouses. Real-life warehouses often have a rectangular layout with few cross-aisles (horizontal lines).
The experiments were performed on an Intel Xeon E5-2440 v2 @ 1.9 GHz processor and 32 GB of RAM. The experiments ran with a memory limit of 8 GB of RAM.

\subsection{Results on rectilinear traveling salesman problem.}
\subsubsection{Pre-processing.}
\label{prepro}
To improve the execution time of the algorithm, we observe that any layout that contains a shortest path $\pi$ of length $l_1(p_1, p_2)$ between each pair of vertices $p_1, p_2 \in P$ is valid to solve the problem. 
Finding the distance preserver (1-spanner) with the minimum number of edges is an NP-complete problem, namely the minimum Manhattan network problem. In practice, if $n$ is not too big ($n \le 1000$), this pre-processing of the graph improves the total execution time. Table~\ref{exper} presents the execution time with and without the computation of the minimum Manhattan network problem with CPLEX 12.5, using a flow formulation (see \cite{nou05}). Notice that since any distance-preserver graph can be used to compute the minimum subtour, we can also apply an approximation algorithm as a pre-process, such as the 2-factor approximation algorithm \cite{nou05} or \cite{guo08} running in $O(n \log n)$.

\subsubsection{Experiments.}
Table~\ref{exper} provides the average and maximum computation time in seconds needed to solve random instances with $n \in \{50, 100, 200\}$ and $h$ varying from 1 to 8. It also reports the maximum number of states obtained on one layer during the computation. We report the results obtained with algorithm \ref{algogo} (column \emph{no pre-proc}) as well as algorithm \ref{algogo} extended with the pre-processing step (column \emph{pre-proc}).
We generated 100 random instances for each configuration \emph{i.e.} for each pair $(n,h)$.
Firstly the algorithm can efficiently handle instances with up to 8 horizontal lines. $h=9$ is not reported since the algorithm runs out of memory. Secondly, the increase of time appears to be roughly linear in practice as $n$ increases for a given $h$. Finally, the maximum number of states matches exactly the values of $|\Omega(h)|$ (see table \ref{tGH}) showing that the worst case is always reached at least on one layer.

\begin{table}[h]
\centering
{\footnotesize
\begin{tabular}{c|c|c|c|c||c|c|c|c||c|c|c|c||c}
\cline{2-13}
                        & \multicolumn{4}{c||}{$n=50$}                                                  & \multicolumn{4}{c||}{$n=100$}                                                 & \multicolumn{4}{c||}{$n=200$}                                                 &                                     \\ \cline{2-14}
                        & \multicolumn{2}{c|}{no pre-proc.} & \multicolumn{2}{c||}{pre-proc.} & \multicolumn{2}{c|}{no pre-proc.} & \multicolumn{2}{c||}{pre-proc.} & \multicolumn{2}{c|}{no pre-proc.} & \multicolumn{2}{c||}{pre-proc.} &   \multicolumn{1}{c|}{max.} \\ \hline
\multicolumn{1}{|c|}{h} & avg.          & max.          & avg.         & max.         & avg.           & max.          & avg.         & max.         & avg.           & max.          & avg.         & max.        & \multicolumn{1}{c|}{states} \\ \hline
\multicolumn{1}{|c|}{1} & 0                  & 0.01              & 0                & 0.01             & 0                  & 0.02              & 0                & 0.01             & 0                  & 0.02              & 0                & 0.02             & \multicolumn{1}{c|}{2}              \\ \hline
\multicolumn{1}{|c|}{2} & 0                  & 0.01              & 0                & 0                & 0                  & 0.04              & 0                & 0.02             & 0                  & 0.05              & 0                & 0.03             & \multicolumn{1}{c|}{6}              \\ \hline
\multicolumn{1}{|c|}{3} & 0                  & 0.01              & 0                & 0                & 0.01               & 0.01              & 0                & 0.02             & 0.01               & 0.02              & 0.01             & 0.06             & \multicolumn{1}{c|}{24}             \\ \hline
\multicolumn{1}{|c|}{4} & 0.02               & 0.04              & 0.01             & 0.01             & 0.04               & 0.06              & 0.02             & 0.03             & 0.08               & 0.11              & 0.04             & 0.06             & \multicolumn{1}{c|}{112}            \\ \hline
\multicolumn{1}{|c|}{5} & 0.12               & 0.18              & 0.05             & 0.08             & 0.27               & 0.33              & 0.12             & 0.18             & 0.51               & 0.65              & 0.27             & 0.4              & \multicolumn{1}{c|}{568}            \\ \hline
\multicolumn{1}{|c|}{6} & 0.76               & 1.09              & 0.26             & 0.42             & 1.72               & 2.52              & 0.78             & 1.23             & 3.56               & 4.84              & 1.78             & 2.73             & \multicolumn{1}{c|}{3032}           \\ \hline
\multicolumn{1}{|c|}{7} & 4.67               & 8.13              & 1.32             & 3.38             & 12.18              & 16.27             & 4.93             & 7.17             & 25.71              & 27.89             & 13.45            & 15.06            & \multicolumn{1}{c|}{16768}          \\ \hline
\multicolumn{1}{|c|}{8} & 42.18              & 72.01             & 7.66             & 19.74            & 114.11             & 136.28            & 50.78            & 69.6             & 242.49             & 269.56            & 116.55     & 131.73           & \multicolumn{1}{c|}{95200}          \\ \hline
\end{tabular}
}
\vspace{0.2cm}
\caption{Results on RTSP. Execution time in seconds and maximum number of states encountered on one layer.}
\label{exper}
\end{table}

\subsection{Results on rectilinear Steiner tree.}
Table~\ref{experRMST} provides the average and maximum computation time in seconds for solving random instances with $n \in \{50, 100, 200\}$ and $h$ varying from 1 to 11. It also reports the maximum number of states obtained on one layer during the computation. No particular pre-processing is applied here. The execution was aborted (due to memory issues) for the cases denoted by "-" in the table.

\begin{table}[h]
\centering
\footnotesize
\begin{tabular}{c|c|c|c|c|c|c|c}
\cline{2-7}
                        & \multicolumn{2}{c|}{$n=50$} & \multicolumn{2}{c|}{$n=100$} & \multicolumn{2}{c|}{$n=200$} &                                  \\ \hline
\multicolumn{1}{|c|}{h} & avg.         & max.         & avg.          & max.         & avg.          & max.         & \multicolumn{1}{c|}{max. states} \\ \hline
\multicolumn{1}{|c|}{1} & 0            & 0.01         & 0             & 0.01         & 0             & 0.01         & \multicolumn{1}{c|}{2}           \\ \hline
\multicolumn{1}{|c|}{2} & 0            & 0.01         & 0             & 0.01         & 0             & 0.01         & \multicolumn{1}{c|}{5}           \\ \hline
\multicolumn{1}{|c|}{3} & 0            & 0.01         & 0             & 0.01         & 0.01          & 0.01         & \multicolumn{1}{c|}{15}          \\ \hline
\multicolumn{1}{|c|}{4} & 0            & 0.01         & 0.01          & 0.01         & 0.02          & 0.03         & \multicolumn{1}{c|}{51}          \\ \hline
\multicolumn{1}{|c|}{5} & 0.01         & 0.02         & 0.04          & 0.06         & 0.09          & 0.12         & \multicolumn{1}{c|}{188}         \\ \hline
\multicolumn{1}{|c|}{6} & 0.03         & 0.05         & 0.15          & 0.21         & 0.36          & 0.5          & \multicolumn{1}{c|}{731}         \\ \hline
\multicolumn{1}{|c|}{7} & 0.11         & 0.19         & 0.58           & 0.98         & 1.55         & 2.35         & \multicolumn{1}{c|}{2950}        \\ \hline
\multicolumn{1}{|c|}{8} & 0.36         & 0.57         & 2.3          & 3.9         & 7.16         & 8.03        & \multicolumn{1}{c|}{12235}       \\ \hline
\multicolumn{1}{|c|}{9} & 1.2         & 2.4        & 10.74         & 13.8        & 33.97        & 55.2        & \multicolumn{1}{c|}{51822}       \\ \hline
\multicolumn{1}{|c|}{10} & 4,46         & 14.05        & 49.63         & 87.13        & -        & -        & \multicolumn{1}{c|}{222616}       \\ \hline
\multicolumn{1}{|c|}{11} & 16,92         & 52.41        & -         & -        & -        & -        & \multicolumn{1}{c|}{771128}       \\ \hline
\end{tabular}
\vspace{0.2cm}
\caption{Results on RST. Execution time in seconds and maximum number of states encountered on one layer.}
\label{experRMST}
\end{table}

\section{Concluding remarks.}
\label{conc}

We introduced a new fixed parameter algorithm for the rectilinear TSP that can efficiently solve instances where the points lie on a few number of horizontal lines. The complexity analysis proves that RTSP can be solved in time $O\left(nh7^h\right)$. Moreover, this algorithm is immediately adapted to solve the rectilinear Steiner tree problem with a $O\left(nh5^h\right)$ runtime and improves over the best known fixed parameter algorithm using the exact same parameter. 

As future work, we are investigating how the algorithm for rectilinear TSP can be used to design very efficient exact methods for the picking problem as well as the joint order batching and picker routing problem in rectangular warehouses \cite{Valle2017}.

\section*{Acknowledgements}
The authors would like to thank V.V. Kruchinin and D.V. Kruchinin for their help with Theorem \ref{theoBnd}, L. Esperet for his explanation of the singularity method and the anonymous reviewers for their valuable comments and suggestions to improve the manuscript.

\bibliography{biblio}
\end{document}